\documentclass[prc,superscriptaddress,a4,nofootinbib]{revtex4}
\usepackage{color}
\usepackage{graphicx}
\usepackage{amssymb}
\usepackage{amsmath,amssymb,bm}
\usepackage{physics}
\usepackage{ulem}

\begin{document}
\title{
Influence of nuclear matter properties on heavy-ion sub-barrier fusion reactions}

\author{Ki-Seok Choi}
\affiliation{School of Liberal Arts and Science, Korea Aerospace University, Goyang 10540, South Korea}

\author{Hana Gil}
\email{khn1219@gmail.com}
\affiliation{Department of Radiological Science, Kangwon National University at Dogye, Samcheok 25945, South Korea}

\author{Kouichi Hagino}
\affiliation{Department of Physics, Kyoto University, Kyoto 606-8502, Japan}
\affiliation{Institute for Liberal Arts and Sciences, Kyoto University, Kyoto 606-8501, Japan}
\affiliation{ 
RIKEN Nishina Center for Accelerator-based Science, RIKEN, Wako 351-0198, Japan}

\author{W. Y. So}
\affiliation{Department of Radiological Science, Kangwon National University at Dogye, Samcheok 25945, South Korea}

\author{Chang Ho Hyun}
\affiliation{Department of Physics Education, Daegu University, Gyeongsan 38453, South Korea}

\author{K. S. Kim}
\affiliation{School of Liberal Arts and Science, Korea Aerospace University, Goyang 10540, South Korea}

\date{\today}

\begin{abstract}
Interactions between two nuclei in nuclear reactions are calculated
by applying the energy density method to nuclear energy density functionals.
We particularly focus on the effects of nuclear matter properties on fusion reactions 
by using the Korea-IBS-Daegu-SKKU (KIDS) density functional. 
Uncertainty of the equation of state in symmetric nuclear matter is accounted for by using different
values of the incompressibility and the effect is explored with the $^{40}$Ca+$^{40}$Ca system.
Though not prominent, a systematic dependence on the incompressibility
can be inferred from the fusion cross sections.
The effect of asymmetric nuclear matter is described by adopting the symmetry energy
parameters that are consistent with the neutron star data, but still have large uncertainties.
In the application to the 
$^{48}$Ca+$^{48}$Ca and
$^{16}$O+$^{208}$Pb systems, we observe a systematic dependence
on the stiffness of the symmetry energy.
We also consider the uncertainty of the effective mass of the nucleon in
nuclear medium.
We employ models that have isoscalar and isovector effective masses varying from
minimum to maximum values in the allowed ranges.
Contrary to incompressibility and symmetry energy, fusion cross sections are
insensitive to the isoscalar and isovector effective masses.
Without any 
phenomenological adjustment or modification of the model, fusion processes are
described accurately by applying the interaction potentials derived from nuclear 
structure models. 
\end{abstract}
\maketitle

\section{Introduction}

A unified description of nuclear structure and nuclear reaction is an important  
goal in the nuclear theory.
The unification is especially important in the reliable prediction of reactions
for which experimental measurement is strongly limited, such as reactions of neutron-rich nuclei as well as fusion 
reactions at astrophysical energies.
For that purpose,
a consistent application of the interaction potential is one of the most important ingredients in deriving the interactions
between the two nuclei that participate in the 
reactions from a given nuclear structure model.

The inter-nucleus potential 
plays critical roles in various nuclear reactions:
it determines the height and the position of the strong absorption barrier, 
that is, the Coulomb barrier,
so consequently has a great impact on the fusion probability and
elastic scattering cross sections.
Microscopically, the inter-nucleus potential can be calculated with e.g., the double folding 
procedure \cite{satchler1979,brandan1997,khoa2000}, the mapping of time-dependent Hartree-Fock (TDHF) 
on a classical Hamiltonian \cite{washiyama2008}, 
the density constrained TDHF method \cite{umar2006,umar2014,umar2009}, 
or the energy density method \cite{npa1975,npa1976}. 
In these approaches, the potential 
is calculated basically from
the interaction of one nucleon in the projectile nucleus with another nucleon in the target one,
and the interactions are summed over the nucleons in projectile and target nuclei. 
Therefore the most fundamental quantities that constitute the nucleus-nucleus potential
are the interaction of nucleons in nuclear medium and density distributions of the
nucleon in projectile and target nuclei.

A series of recent works have explored the effect of nuclear matter equation of state (EoS) on the fusion process.
Reference \cite{epja2023} considered the impact of density dependence of the symmetry energy
on astrophysical $S$-factor for heavy-ion fusion reactions.
References \cite{prc2024} and \cite{npa2025} focused on the correlation between
incompressibility of symmetric nuclear matter and fusion observables.
The former work used relativistic mean field 
models to obtain the wave functions of
target and projectile nuclei, while the latter works adopted Skyrme forces.
All of the three works commonly estimated the nucleus-nucleus potential 
using the double folding
procedure, in which nucleon-nucleon interactions are described in terms of various
versions of M3Y potential.
It has been demonstrated that the fusion reactions are closely related to
incompressibility and symmetry energy.

In this work, we calculate the nucleus-nucleus potential by using the
energy density method (EDM), in which the interaction potential is obtained
as a difference between the energy of projectile-target composite system
and the energies of isolated projectile and target nuclei \cite{npa1975,npa1976,brueckner1968,ngo1975,dobrowolski2003,liu2006,zamrun2006,shodsi2026}.
In the density functional theory (DFT), the energy of a nucleus is a functional 
of the density distribution of the nucleus, which is determined by solving the
Hartree-Fock (or the Kohn-Sham) equation with a given single-particle mean field potential.
The solution of the Hartree-Fock (the Kohn-Sham) equation is inserted into the energy
density functional (EDF) from which single-particle potential is calculated 
to obtain the energy of a nucleus, so the interaction potential is 
determined self-consistently in the framework of DFT.

In this paper, we employ 
Korea-IBS-Daegu-SKKU (KIDS) density functional \cite{prc2018} for the EDF.
KIDS functional was initiated with the aim of a unified description of finite nuclei and infinite nuclear matter.
It has been shown that KIDS functional satisfies the nuclear matter constraints
from heavy-ion collisions, neutron star mass-radius data and the tidal deformability
from the observation of gravitational waves \cite{epja2020}.
At the same time, it reproduces basic properties of nuclei such as binding energies
and charge radii at the accuracy of 1~\% or less deviation from experimental data \cite{prc2019a, prc2019b}.
A great advantage of the KIDS model is that uncertainties in the EoS
of nuclear matter and properties of nuclei can be controlled and examined
systematically without spoiling the unified description of nuclei and nuclear matter.
The functional has also proven to be applicable to $\alpha$-decay of heavy isotopes \cite{prc2025}, quasielastic exclusive \cite{prc2021} and inclusive electron scattering \cite{prc2022},
and neutrino-nucleus scattering \cite{plb2022} without any complementary adjustment
of the model to decay or scattering processes.
Now it is timely to test the applicability of KIDS functional in the low-energy
heavy ion reactions.

Utilizing the advantage of the KIDS model,
we examine the effect of uncertainties from nuclear matter and finite nuclei 
in the low-energy fusion reactions.
More specifically, ambiguities in the density dependence of the nuclear matter EoS are
considered by adopting KIDS models \cite{npsm2021}.
The models are different or similar in the incompressibility of symmetric nuclear matter, $K_0$,
the symmetry energy at saturation, $J$, and the slope of the symmetry energy, $L$.
In order to discriminate the effect of each parameter, we consider the fusion processes of 
$^{40}$Ca+$^{40}$Ca, $^{48}$Ca+$^{48}$Ca 
and $^{16}$O+$^{208}$Pb systems.
In $^{40}$Ca+$^{40}$Ca, the number of protons is equal to that of neutrons, so the effect of symmetry
energy is maximally suppressed.
Consequently, the system is expected to exhibit the role of $K_0$.
In the latter two systems, 
because $^{48}$Ca and $^{208}$Pb are asymmetric in the number of protons and neutrons,
the effect of the symmetry energy
can be manifested in these systems.

In addition to the uncertainties of the nuclear matter EoS, we also consider the 
sensitivity to the in-medium effective mass of the nucleon on the fusion reaction.
In the Skyrme EDF, coefficients that contribute to the effective mass also participate 
in the terms relevant to the derivative of density, 
that is most significant in the surface region of the nuclei.
Because fusion reaction is sensitive to the potential barrier which is formed
in the surface region,
one may expect that the effective mass can have certain effects on the fusion process.

We organize the paper in the following order.
In Sec. II, we present two basic building blocks of the work, that is, 
the nuclear EDF and 
the EDM for nucleus-nucleus potential.
In Sec. III, we show the results of the dependence on EoS with
density profile of the considered nuclei, 
nucleus-nucleus potentials for $^{40}$Ca+$^{40}$Ca,
$^{48}$Ca+$^{48}$Ca and $^{16}$O+$^{208}$Pb systems,
their fusion cross sections and barrier distributions obtained from the four EDF models.
Impact of the effective mass on fusion is presented with the results
of potential and fusion cross sections.
The results are also compared with available experimental data.
We summarize the work in Sec. IV.

\section{Formalism}
\subsection{KIDS functional}

The starting point of the KIDS model is the energy per particle in nuclear matter expanded in terms of
$\rho^{1/3}$ as \cite{prc2018}
\begin{equation}
{\cal E}(\rho, \delta) = \sum^n_{i=0} (\alpha_i + \beta_i \delta^2) \rho^{1+i/3},
\label{eq:ekids}
\end{equation}
where $\rho=\rho_n+\rho_p$ and $\delta = (\rho_n-\rho_p)/\rho$.
Here, $\rho_n$ and $\rho_p$ are the densities of the neutron and the proton, respectively.
The coefficients $\alpha_i$ and $\beta_i$ are determined
to reproduce nuclear matter properties in the first step.
In order to apply the model to finite nuclei in the second step, 
we transform the energy functional
in Eq. (\ref{eq:ekids}) to the form of Skyrme functional.
In the course of the transformation, $\alpha_i$ and $\beta_i$ determined by 
the nuclear matter EoS are kept unchanged and 
terms that account for
the derivative of density
and spin-orbit couplings are added.
Coefficients of these additional terms are determined to reproduce nuclear data
accurately but they do not affect the nuclear matter properties.
Because the nuclear matter and finite nuclei are adjusted in separate steps,
one can describe both systems accurately in a single framework.

A conventional way to express energy per particle in homogeneous nuclear matter is
\begin{eqnarray}
{\cal E}(\rho, \delta) &=& E(\rho) + S(\rho) \delta^2 + O(\delta^4), 
\label{eq:E}
\\
E(\rho) &=& E_B + \frac{1}{2} K_0 x^2 + \frac{1}{6} Q_0 x^3 + \cdots, 
\label{eq:E-rho}
\\
S(\rho) &=& J + L x + \frac{1}{2} K_{\rm sym} x^2 + \frac{1}{6} Q_{\rm sym} x^3
+ \cdots,
\label{eq:S}
\end{eqnarray}
where $x = (\rho-\rho_0)/3\rho_0$ and $\rho_0$ is the nuclear saturation density.
The coefficients $K_0$, $L$ and $K_{\rm sym}$ 
are key parameters to determine the behavior of energy
below and above $\rho_0$ in nuclei and neutron stars.

Density dependence of the nuclear EoS constitutes an
essential issue in both finite nuclei and nuclear matter systems.
For instance, the slope parameter of the symmetry energy $L$ has strong correlations 
with radius and tidal deformability 
of the canonical neutron star and neutron skin thickness of neutron-rich nuclei.
Cross sections of the nucleus-nucleus reactions at low energies are known to be
sensitive to the density profile in the surface region of the projectile and
target nuclei,
where density decreases below $\rho_0$.
These observations lead to a conjecture that the density dependence of the EoS
can have non-negligible effects in nuclear reactions.
To accommodate the uncertainties of the EoS in the nuclear reactions,
we derive the nucleus-nucleus potential by applying EDM
to nuclear EDF models and explore the role of the parameters that 
determine the density dependence of nuclear EoS in nuclear fusion reactions.
We consider four models that are distinguished in the 
incompressibility of the symmetric nuclear matter and
the density dependence of the symmetry energy.
The parameters in the EoS of the models are summarized in 
Tab.~\ref{tab:eos}, in which
$\mu_{\rm IS}$ and $\mu_{\rm IV}$ 
respectively denote the isoscalar 
and isovector effective masses at saturation density 
divided by the nucleon mass in free space, that is,
\begin{eqnarray}
\mu_{\rm IS} = \frac{m^*_{\rm IS}(\rho=\rho_0)}{m}, 
\,\,\,\,
\mu_{\rm IV} = \frac{m^*_{\rm IV}(\rho=\rho_0)}{m}.
\label{eq:muISIV}
\end{eqnarray}

\begin{table}[t]
\caption{Nuclear matter properties of models employed in the work.
See Eqs. (\ref{eq:E}), (\ref{eq:E-rho}), and (\ref{eq:S}) for the definitions of the nuclear matter 
quantities. 
The unit of the saturation density, $\rho_0$ is fm$^{-3}$ and it is MeV for 
$E_B$, $K_0$, $J$, $L$ and $K_{\rm sym}$.
Values are taken 
from Ref. \cite{npsm2021} for the KIDS-A, B, C, D models. 
Definitions of
$\mu_{\rm IS}$ and $\mu_{\rm IV}$ are given by Eq.~(\ref{eq:muISIV}).
}
\begin{center}
\begin{ruledtabular}
\begin{tabular}{ccccccccc}
 \,\,\, Model \,\,\, &\,\,\,  $\rho_0$\,\,\,  &\,\,\,  $E_B$\,\,\,  & 
\,\,\,  $K_0$\,\,\,  &\,\,\,  $J$\,\,\,  & 
\,\,\, $L$\,\,\,  &\,\,\,  $K_{\rm sym}$\,\,\,
& \,\,\, $\mu_{\rm IS}$ \,\,\, & \,\,\, $\mu_{\rm IV}$ \,\,\,
\\ \hline
KIDS-A   & 0.160 & 16.0 & 230 & 33 & 66 & $-139.5$ & $1.005$ & $0.814$ \\
KIDS-B   & 0.160 & 16.0 & 240 & 32 & 58 & $-162.1$ & $0.997$ & $0.812$ \\
KIDS-C   & 0.160 & 16.0 & 250 & 31 & 58 & $-91.5$ & $0.983$ & $0.805$ \\
KIDS-D   & 0.160 & 16.0 & 260 & 30 & 47 & $-134.5$ & $0.970$ & $0.799$ \\ \end{tabular}
\end{ruledtabular}
\end{center}
\label{tab:eos}
\end{table}

\subsection{Nucleus-nucleus potential}
\label{sec:formalism-nucleus-nucleus-potential}

\begin{figure}
\begin{center}
\includegraphics[width=0.7\textwidth]{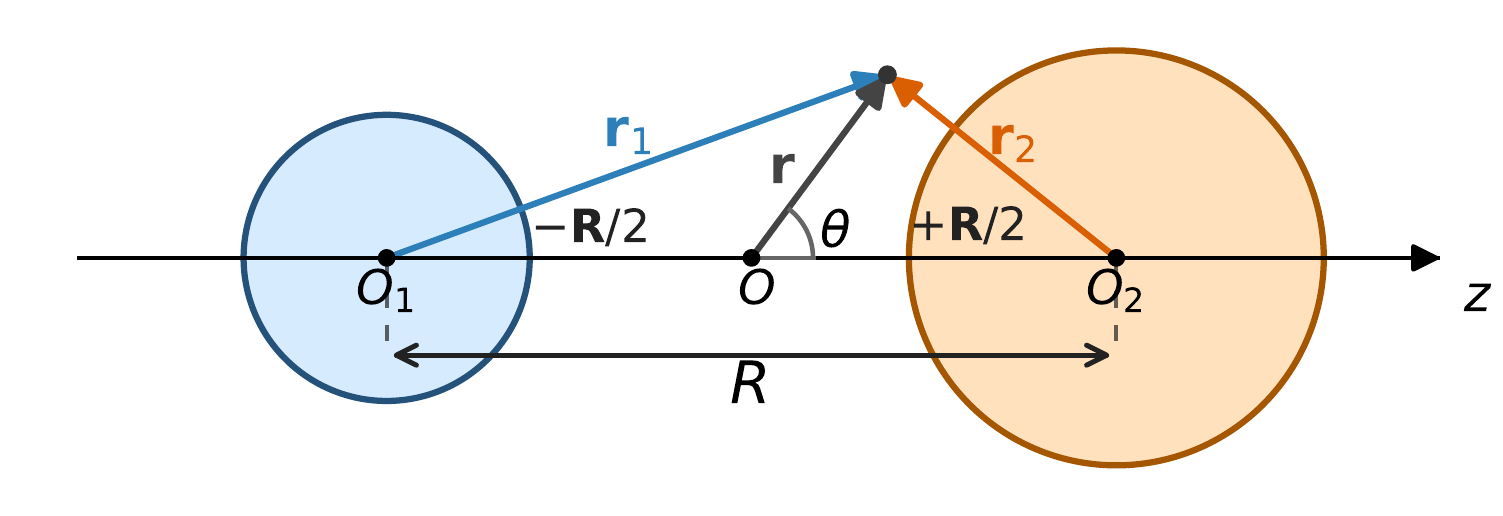}
\end{center}
\caption{Definitions of the coordinates in the evaluation 
of a nuclear potential $V_N(R)$.}
\label{fig:coordinate}
\end{figure}

The internuclear potential is calculated with the frozen-density EDM.  
In this approach the two nuclei are assumed
to keep their ground-state density profiles during the collision. 
The 
interaction energy is then obtained from the change of the total energy density
when the two nuclei overlap.  
For two nuclei separated by the
center-of-mass distance \(R\), the nuclear part of the potential is written as
\begin{equation}
  V_N(R)
  =
  E_T(R)-E_1(R)-E_2(R) ,
  \label{eq:vn-edm}
\end{equation}
where
\begin{align}
  E_T(R)
  &=
    \int d^3 r\,
  {\cal H} \left (
    \rho_{1n}+\rho_{2n},
    \rho_{1p}+\rho_{2p}
  \right),
  \label{eq:edm-total-energy}
  \\
  E_i(R)
  &=
  \int d^3 r\,
  {\cal H}
  \left(
    \rho_{in},
    \rho_{ip}
  \right),
  \qquad i=1,2 .
  \label{eq:edm-isolated-energy}
\end{align}
Here \({\cal H}\) is the nuclear Hamiltonian density, i.e. the local energy
density as a functional of the neutron and the proton densities.  The quantities
\(\rho_{in}\) and \(\rho_{ip}\) denote the neutron and the proton densities of the 
nucleus \(i\), respectively.  The densities in
Eq.~\eqref{eq:edm-total-energy} are evaluated in the frozen-density
approximation, so the total density is formed by adding the two isolated
nuclear densities without rearranging them self-consistently.  
The coordinate systems are defined as depicted in Fig.\,\ref{fig:coordinate}.
For a spherical integration coordinate \(r\) and
\(\mu=\cos\theta\), $r_1$ and $r_2$ are given by
\begin{equation}
  r_1 =
  \left(r^2+\frac{R^2}{4}+rR\mu\right)^{1/2},
  \,\,\,
  r_2 =
  \left(r^2+\frac{R^2}{4}-rR\mu\right)^{1/2}, 
  \label{eq:r1-r2}
\end{equation}
respectively. The 
variables of the densities 
in Eqs.~(\ref{eq:edm-total-energy}, \ref{eq:edm-isolated-energy}) are
$\rho_{iq} = \rho_{iq}(r_i)$ ($q=n, p$).

In this work, \({\cal H}\) is taken from the KIDS EDF mapped onto a Skyrme-type form.  
The corresponding KIDS Hamiltonian density used in the present calculation is
\begin{align}
  {\cal H}
  =&\,
  \frac{\hbar^2}{2m} \tau 
  +\frac{3}{8}t_0\rho^2
  -\frac{1}{8}(t_0+2y_0)(\rho_n-\rho_p)^2
  \nonumber\\
  &+
  \sum_{n=1}^{3}
  \left[
    \frac{1}{16}t_{3n}\rho^{2+n/3}
    -
    \frac{1}{48}(t_{3n}+2y_{3n})
    \rho^{n/3}
    (\rho_n-\rho_p)^2
  \right]
  \nonumber\\
  &+
  \frac{1}{8}
  (2t_1+y_1+2t_2+y_2)\rho\tau
  -
  \frac{1}{8}
  \left[
    (t_1+2y_1)-(t_2+2y_2)
  \right]
  \sum_{q=n,p}\rho_q\tau_q
  \nonumber\\
  &+
  \frac{1}{64}
  (9t_1-5t_2-4y_2)
  (\nabla\rho)^2
  -
  \frac{1}{64}
  (3t_1+6y_1+t_2+2y_2)
  (\nabla \rho_n - \nabla \rho_p)^2
  \nonumber\\
  &+
  \frac{1}{2}W_0
  \left[
    {\bf J}\cdot\nabla\rho
    +
    \sum_{q=n,p}
    {\bf J}_q\cdot\nabla\rho_q
  \right],
  \label{eq:kids-hamiltonian-density}
\end{align}
where the local densities entering the EDF are
\begin{equation}
  \rho=\rho_n+\rho_p,
  \qquad
  \tau=\tau_n+\tau_p ,
  \qquad
  {\bf J}={\bf J}_n+{\bf J}_p .
  \label{eq:density-definitions}
\end{equation}
Here $m$ is the mass of the nucleon in free space,
\(\tau_q\) is the kinetic energy density, 
and \({\bf J}_q\) is the spin-current density for isospin \(q=n,p\). 
Equation~\eqref{eq:kids-hamiltonian-density} is the energy-density form of
the KIDS functional.  
The terms in the third line of Eq. (\ref{eq:kids-hamiltonian-density}) are included 
in the kinetic term in the first line to define the isoscalar and isovector effective masses
\begin{eqnarray}
\frac{m^*_{\rm IS}}{m} &=& \left( 1+ \frac{m}{8\hbar^2} \rho \Theta_s \right)^{-1},
\label{eq:iseffmass}
\\
\frac{m^*_{\rm IV}}{m} &=& \left( 1 + \frac{m}{4\hbar^2} \rho \Theta_v \right)^{-1},
\label{eq:iveffmass}
\end{eqnarray}
where $\Theta_s = 3t_1 + 5 t_2 + 4y_2$ and $\Theta_v = 2t_1 + y_1 + 2t_2 +y_2$.
Numerical values of the coefficients of KIDS functional in Eq. (\ref{eq:kids-hamiltonian-density}) are summarized in Tab. \ref{tab:parameter}.

\begin{table}[t]
\caption{
Parameters for the KIDS functional in the form of Skyrme force.
Units of the coefficients are MeV\,fm$^3$ for $t_0$ and $y_0$,
MeV\,fm$^4$ for $t_{31}$ and $y_{31}$,
MeV\,fm$^5$ for $t_1$, $y_1$, $t_2$, $y_2$, $t_{32}$, $y_{32}$, and  $W_0$,
and MeV\,fm$^6$ for $y_{33}$. 
$y_1$, $y_2$ and $t_{33}$ are identically zero in the
KIDS functionals.}
\begin{center}
\begin{ruledtabular}
\begin{tabular}{ccccc}
Parameter & KIDS-A & KIDS-B & KIDS-C & KIDS-D \\ \hline
$t_0$ & $-1855.377$ & $-1772.044$ & $-1688.710$ & $-1605.377$ \\
$y_0$ & $ 2302.760 $ & $2057.283$ & $751.561 $ & $543.801 $ \\
$t_1$ & $300.737 $ & $296.306$ & $295.097 $ & $293.565 $ \\
$t_2$ & $-182.308 $ & $-176.616$ & $-169.851 $ & $-163.281 $ \\
$t_{31}$ & $14058.746 $ & 12216.730 & $10374.714 $ & $8532.698 $ \\
$y_{31}$ & $-76808.423 $ & $-70716.593$ & $-33487.909 $ & $-29128.971 $ \\
$t_{32}$ & $-1020.642 $ & 620.997 & $2208.265 $ & $3802.565 $ \\
$y_{32}$ & $127228.272 $ & 118923.207 & $64791.066 $ & $59087.582 $ \\
$y_{33}$ & $-75362.002 $ & $-70290.560$ & $-46589.511 $ & $-41756.516 $ \\
$W_0$ & $98.683 $ & 98.292 & $101.306 $ & $101.664 $ \\
\end{tabular}
\end{ruledtabular}
\end{center}
\label{tab:parameter}
\end{table}

In EDM,
the kinetic energy density is evaluated in the semiclassical extended
Thomas--Fermi (ETF) approximation.  This approximation expresses \(\tau_q\)
locally in terms of the density, density gradients, momentum-dependent
form-factor corrections, and spin-orbit corrections instead of solving explicit
single-particle wave functions \cite{dobrowolski2003,liu2006,shodsi2026} (See also Refs. \cite{wu2022,wu2025,colo2023,wu2026} for recent 
developments of the orbital-free density functional theory (OF-DFT) for nuclear systems).  
For \(q=n,p\), the kinetic energy density then reads,
\begin{align}
  \tau_q
  =&\,
  \frac{3}{5}(3\pi^2)^{2/3}\rho_q^{5/3}
  +
  \frac{(\nabla\rho_q)^2}{36\rho_q}
  +
  \frac{\nabla^2\rho_q}{3}
  +
  \frac{\nabla\rho_q\cdot\nabla f_q+\rho_q\nabla^2 f_q}{6f_q}
  \nonumber\\
  &-
  \frac{\rho_q}{12}
  \left(
    \frac{\nabla f_q}{f_q}
  \right)^2
  +
  \frac{1}{2}\rho_q
  \left[
    \frac{2m}{\hbar^2}
    \frac{W_0}{2}
    \frac{\nabla(\rho+\rho_q)}{f_q}
  \right]^2 ,
  \label{eq:etf-tau}
\end{align}
where 
\begin{align}
  f_q
  =\,
  1+
  \frac{2m}{\hbar^2}
  \frac{1}{8}
  (2t_1+y_1+2t_2+y_2)\rho
  - \frac{2m}{\hbar^2}
  \frac{1}{8}
  \left[
    (t_1+2y_1)-(t_2+2y_2)
  \right]\rho_q .
  \label{eq:momentum-dependent-form-factor}
\end{align}
The dimensionless quantity \(f_q\) is the form factor generated by the
momentum-dependent terms of the Skyrme-type KIDS functional.  In the present
calculation this \(f_q\) is used directly in the gradient expansion of
\(\tau_q\).  The ETF spin-current density is taken as
\begin{equation}
  {\bf J}_q
  =
  -\frac{2m}{\hbar^2}
  \frac{W_0}{2}
  \frac{\rho_q}{f_q}
  \nabla(\rho+\rho_q).
  \label{eq:spin-current-etf}
\end{equation}
The spin-current density \({\bf J}_q\) describes the local spin-orbit
polarization induced by the density gradient. 

\section{Results}

\subsection{Ground-state density inputs}

The ground-state proton and neutron density distributions of
$^{16}\mathrm{O}$, $^{40}\mathrm{Ca}$, 
$^{48}$Ca and $^{208}\mathrm{Pb}$ 
serve as input to the frozen-density EDM. 
It is therefore useful to examine their dependence on the EoS before discussing the corresponding
nucleus-nucleus potentials.

\begin{figure}[h]
  \centering

\includegraphics[width=0.46\textwidth]{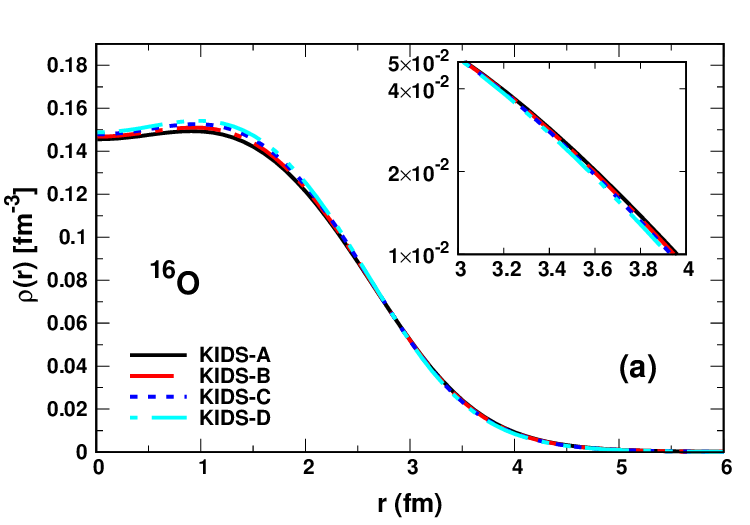}
\includegraphics[width=0.46\textwidth]{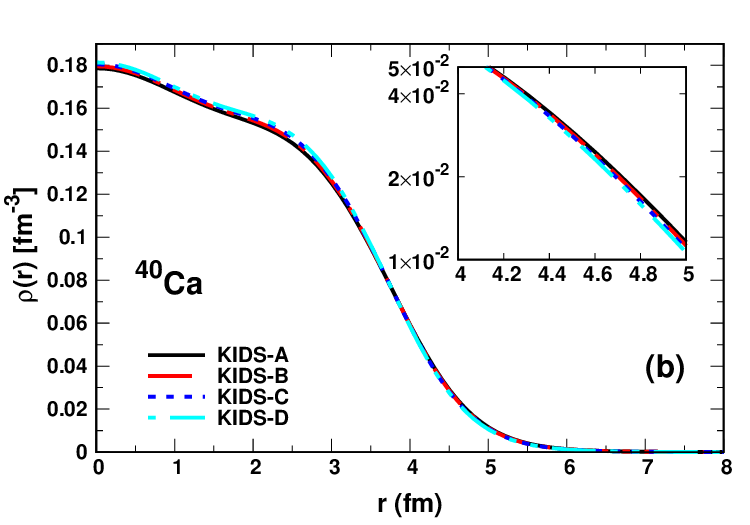}
\includegraphics[width=0.46\textwidth]{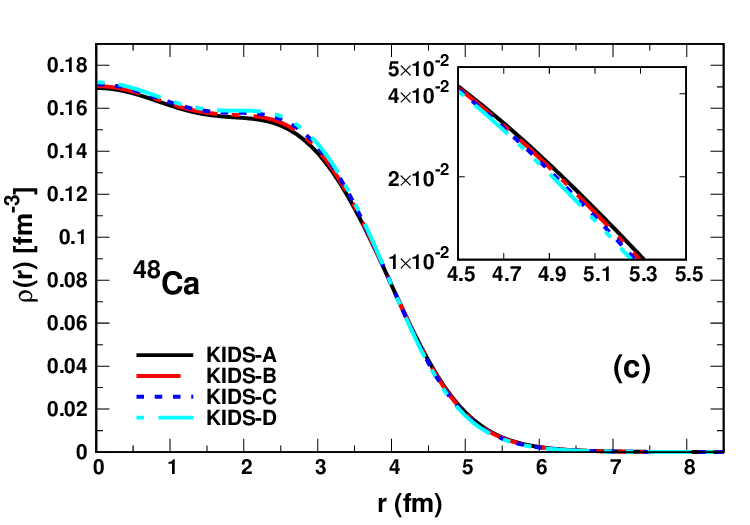}
\includegraphics[width=0.46\textwidth]{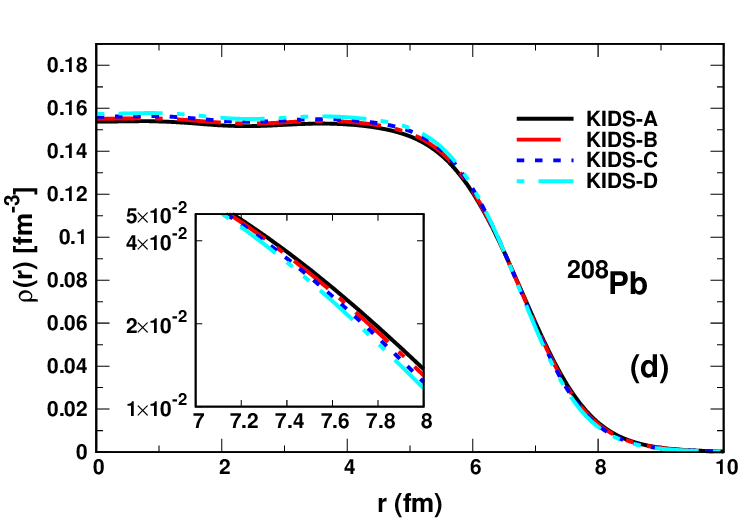}

  \caption{
    Ground-state matter density distributions used as frozen-density
    inputs for the EDM calculation:
    (a) $^{16}\mathrm{O}$, (b) $^{40}\mathrm{Ca}$, 
    (c) $^{48}\mathrm{Ca}$, and (d) $^{208}\mathrm{Pb}$. 
    The curves compare KIDS-A, KIDS-B, KIDS-C and KIDS-D. 
    The surface region is very similar among the
    functionals, while discernible 
    model-dependent differences remain mainly
    in the nuclear interior. The inset shows an enlarged view of the 
    surface region on a logarithmic scale.}
  \label{fig:results-densities}
\end{figure}

Figure~\ref{fig:results-densities} compares the matter
density distributions for the four nuclei used in the present
calculations. 
These are calculated 
by implementing the KIDS-A, KIDS-B, KIDS-C and KIDS-D EDFs in the HFBTHO code \cite{hfbtho}.
To this end, the Coulomb potential is also included in the single-particle mean field equation in the form of 
\begin{eqnarray}
U_{\rm Coul}(\mathbf{r}) = \frac{e^2}{2} \int 
\frac{\rho_C(\mathbf{r'}) d^3 r'}{|\mathbf{r}-\mathbf{r'}|}
-\frac{e^2}{2} \left( \frac{3}{\pi} \right)^{1/3} \rho_C^{1/3}(\mathbf{r}),   
\end{eqnarray}
where the first and second terms on the right hand side represent the
direct and exchange terms, repectively, and $\rho_C$ denotes
the charge density.
For $^{16}\mathrm{O}$, 
small deviations are distinguishable in the central region, where the KIDS parametrizations give slightly different interior densities.
In the central region, magnitude of density is in the order
KIDS-A $<$ KIDS-B $<$ KIDS-C $<$ KIDS-D.
Since $^{16}$O is a symmetric nucleus, i.e. 
$\rho_n \sim \rho_p$,
contribution of the symmetry energy is suppressed because $\delta \sim 0$.
Then the difference between the models can stem from $K_0$.
Because the ordering of matter density in the core follows exactly that
of the incompressibility, one can deduce that $K_0$ mainly causes the dependence on EoS in the symmetric nuclei.

A similar pattern is found for $^{40}\mathrm{Ca}$. 
The magnitude of the density in the central region is in the same order 
as $^{16}$O, so the result supports the conjecture that the density distribution is
sensitive to $K_0$, and that a larger $K_0$ value produces higher 
density in the core.
The same dependence on $K_0$ is observed with
existing Skyrme force models \cite{npa2026}.
If the core is less dense, deficiency of the density in the core will be 
compensated by enhancement of density in the surface region.
Insets in the figure magnify the density of each nucleus 
in the surface region in the logarithmic scale. 
Though the lines are very close to each other, it is clear that the
magnitude of the density is in the reverse order of the density in the core.
Because sub-barrier fusion is governed mainly by
the overlap of the nuclear surfaces, 
a small difference in the surface can lead to a strong direct effect
on the fusion dynamics.

Let us next discuss nuclei with $\delta\neq0$.
The trend observed in $^{16}$O and $^{40}$Ca works in $^{48}$Ca also.
A notable difference from the former two symmetric nuclei 
is in the inset of $^{48}$Ca, while
the lines are very close to each other in the insets of $^{16}$O
and $^{40}$Ca. 
The intervals between the curves become more
spacious in $^{48}$Ca. It is well known that the neutron skin
thickness is strongly correlated with the slope parameter $L$
\cite{ijmpe2022}: a larger $L$ produces a thicker neutron skin.
More roomy space between curves in $^{48}$Ca than 
$^{16}$O and $^{40}$Ca indicates that the symmetry energy 
comes into play in the density of surface region for the
neutron-rich nuclei.

For $^{208}\mathrm{Pb}$, all the four EDFs again predict the same ordering of density in the central region.
However, because $^{208}$Pb is a heavy and
highly asymmetric nucleus, 
the role of symmetry energy becomes more transparent
in the density distribution.
Difference between the models in the 
surface of $^{208}$Pb is more significant than
$^{48}$Ca as can be seen in the inset.
This enhancement of the model dependence in $^{208}$Pb 
may primarily arise from
the large $\delta$ value of $^{208}$Pb and 
different $L$ values in each model.
Consequently, differences in the calculated nuclear potentials and fusion
cross sections are expected to originate from the EDF in the overlap
region where the effect of EoS appears systematically
in the isolated ground-state density profiles.

\subsection{Nuclear potentials}

\begin{figure}[tbp]
  \centering
\includegraphics[width=0.5\textwidth]{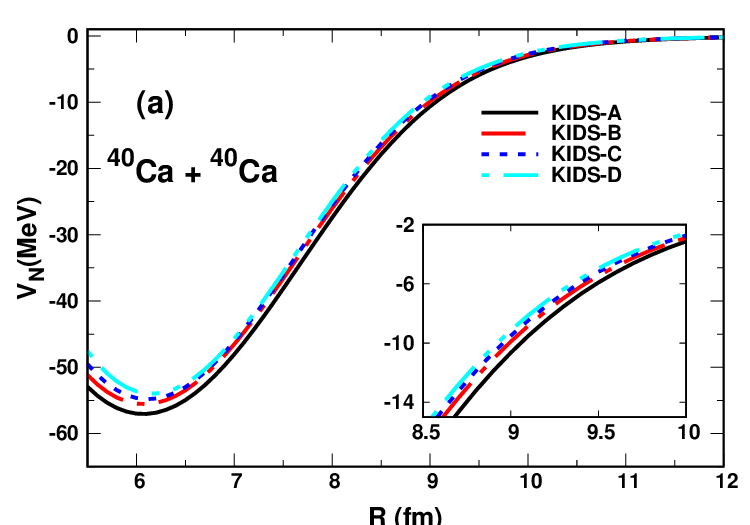}
\includegraphics[width=0.5\textwidth]{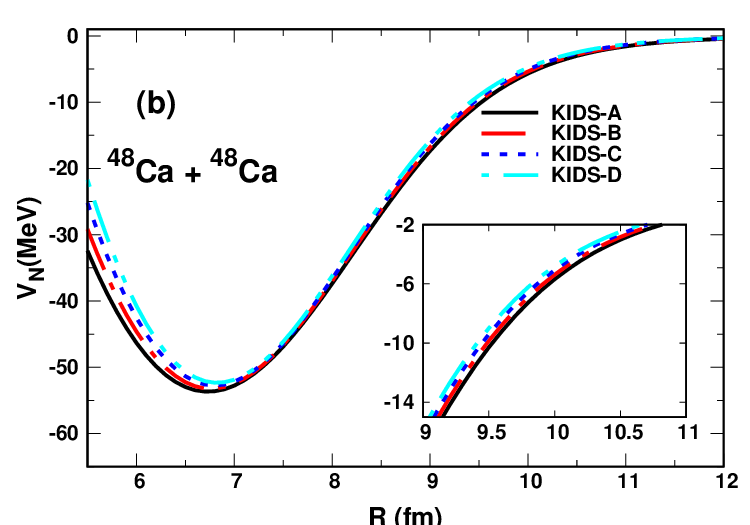}
\includegraphics[width=0.5\textwidth]{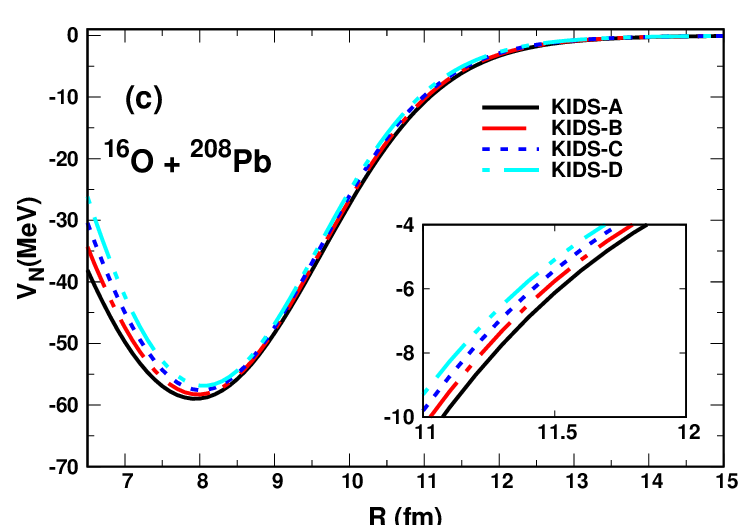}
  \caption{
    Nuclear potentials $V_N(R)$ for
    (a) $^{40}\mathrm{Ca}+{}^{40}\mathrm{Ca}$, 
    (b) $^{48}\mathrm{Ca}+{}^{48}\mathrm{Ca}$, 
    and (c) $^{16}\mathrm{O}+{}^{208}\mathrm{Pb}$ obtained from the frozen-density
    EDM potentials. 
    The curves show the results for KIDS-A, B, C, and D 
    with the line styles indicated in each panel.
    The inset in each panel enlarges the surface region relevant to the
    fusion barrier.
  }
  \label{fig:results-potentials}
\end{figure}

Figure~\ref{fig:results-potentials} displays the nuclear potentials for
(a) $^{40}\mathrm{Ca}+{}^{40}\mathrm{Ca}$,
(b) $^{48}\mathrm{Ca}+{}^{48}\mathrm{Ca}$, and
(c) $^{16}\mathrm{O}+{}^{208}\mathrm{Pb}$, obtained with the
frozen-density EDM using the KIDS-A, KIDS-B, KIDS-C
and KIDS-D EDFs. 
In each panel, the inset enlarges the outer surface region, 
where the functional dependence becomes 
relevant to the formation of the fusion barrier.
The resulting nucleus-nucleus potentials exhibit behavior 
consistent with the density profile of each model.

For $^{40}\mathrm{Ca}+{}^{40}\mathrm{Ca}$, 
the magnitude of the potential follows the same order of density
in the surface region, so we have the strongest attraction with KIDS-A
and the weakest with KIDS-D.
The difference between models reaches several MeV near the potential pocket
and remains visible in the surface region up to $R\simeq 10$ fm. 
According to the density distribution of $^{40}$Ca, the core of the nucleus
extends to $r \simeq 2.5$ fm and the density decreases at a substantial rate 
up to $r \simeq 5$ fm in the surface region, after which tail-like behavior follows.
If we assume $R=r_1 + r_2$, $R=10$ fm is the distance at which 
the surfaces of the two nuclei ($r_1=r_2=5$ fm) start to touch each other.
Therefore, the result implies that the model dependence becomes apparent 
at distances where surfaces overlap.
The inset in Fig.~\ref{fig:results-potentials} (a) compares the potential
at distances near the peak of the Coulomb barrier.
Since the barrier height, radius, and curvature 
summarized in Tab.~\ref{tab:ws-fit-parameters} 
are determined by the
potential in this region, even moderate changes can alter the barrier
penetrability and, consequently, the fusion cross section. 
The EoS dependence is more clearly expressed in the nuclear potential than
in the isolated ground-state densities.

The general trend is similar for
both $^{48}$Ca+$^{48}$Ca and 
$^{16}\mathrm{O}+{}^{208}\mathrm{Pb}$.
The densities of $^{16}$O, $^{48}$Ca,
and $^{208}$Pb are highest for KIDS-A and
lowest for KIDS-D in the surface region, and the potential follows the
same ordering in magnitude.
We observed that symmetry energy promotes stronger model dependence 
in the density profile of $^{48}$Ca and $^{208}$Pb.
Such an effect is also reflected in the potential.
For instance, 
at distances close to peak position, 
the largest gap between the potentials is
0.79 MeV at $R=9.5$ fm for $^{40}$Ca+$^{40}$Ca, but increases to 1.04 MeV
at $R=11.5$ fm for $^{16}$O+$^{208}$Pb.
Enhanced differences among the models can noticeably
modify the tunneling probability because sub-barrier fusion depends
exponentially on the barrier properties, making it sensitive to
changes in the surface potential. 
As in the symmetric system, the influence of the EoS  
appears more evident through the overlap interaction rather than
through the static density distributions.

\begin{table}[t]
  \centering
  \caption{
    Parameters for a Woods-Saxon fit for the calculated nuclear potentials.
    The potential is given by the Woods-Saxon form as in 
    Eq.~\eqref{eq:ws-fit-results}, with
    $R_0=r_0(A_1^{1/3}+A_2^{1/3})$. The parameters $V_0$, $r_0$,
    and $a$ are given in MeV, fm and fm, respectively. The barrier
    quantities $V_b$, $R_b$, and $\hbar\omega$ are the uncoupled
    barrier height, radius, and curvature in units of MeV, fm
    and MeV, respectively.
  }
  \label{tab:ws-fit-parameters}
  \begin{center}
\begin{ruledtabular}
  \begin{tabular}{llcccccc}
    System & Model & $V_0$ (MeV) & $r_0$ (fm) & $a$ (fm)& $V_b$ (MeV) & $R_b$ (fm) & $\hbar\omega$ (MeV)\\
    \hline
    $^{40}\mathrm{Ca}+{}^{40}\mathrm{Ca}$ & KIDS-A & 60.345 & 1.150 & 0.737 & 54.80 & 9.64 & 3.45 \\
     & KIDS-B & 60.663 & 1.145 & 0.732 & 55.04 & 9.60 & 3.48  \\
     & KIDS-C & 60.927 & 1.140 & 0.728 & 55.29 & 9.56 & 3.51 \\
     & KIDS-D & 61.245 & 1.136 & 0.723 & 55.52 & 9.52 & 3.54 \\
        \hline
    $^{48}\mathrm{Ca}+{}^{48}\mathrm{Ca}$ 
    & KIDS-A & 62.067 & 1.142 & 0.741 & 52.04 & 10.20 & 3.03 \\
    & KIDS-B & 63.592 & 1.135 & 0.736 & 52.30 & 10.16 & 3.06 \\ 
    & KIDS-C & 64.735 & 1.130 & 0.726 & 52.57 & 10.12 & 3.11 \\
    & KIDS-D & 66.654 & 1.122 & 0.722 & 52.81 & 10.07 & 3.14 \\
    \hline
    $^{16}\mathrm{O}+{}^{208}\mathrm{Pb}$ & KIDS-A & 68.104 & 1.148 & 0.782 & 76.02 & 11.50 & 4.18 \\
    & KIDS-B & 67.992 & 1.144 & 0.773 & 76.39 & 11.45 & 4.22 \\ 
    & KIDS-C & 67.784 & 1.141 & 0.764 & 76.74 & 11.41 & 4.27 \\
    & KIDS-D & 67.634 & 1.138 & 0.755 & 77.11 & 11.36 & 4.32 \\
\end{tabular}
\end{ruledtabular}
\end{center}
\end{table}

\subsection{Total fusion cross sections}

To carry out coupled-channels calculations, we fit
the calculated nuclear potentials $V_N(R)$
with a Woods-Saxon form, 
\begin{equation}
  V_N^{\mathrm{WS}}(R)
  =
  -\frac{V_0}
  {1+\exp\left[(R-R_0)/a\right]} ,
  \label{eq:ws-fit-results}
\end{equation}
where $V_0$, $R_0$, and $a$ are the fitted depth, radius, and
diffuseness parameters, respectively. 
The radius $R_0$ is often parameterized as,
\begin{equation}
  R_0 = r_0\left(A_1^{1/3}+A_2^{1/3}\right),
  \label{eq:ws-radius-convention-results}
\end{equation}
where $A_1$ and $A_2$ are the mass number of each of the colliding nuclei. 
The fitted parameters are listed in Tab.~\ref{tab:ws-fit-parameters}. 
The same table also lists the
uncoupled barrier height $V_b$, the position $R_b$, and 
the curvature $\hbar \omega$ at the peak of the inter-nucleus potential evaluated with each model. 
To evaluate these quantities, we have added the Coulomb potential with $V_C(R)=Z_1Z_2e^2/R$, 
where $Z_1$ and $Z_2$ are the proton numbers of the colliding nuclei.
These quantities provide a compact
way to compare how the model dependence of the EDM potential is
translated into the barrier parameters relevant to fusion cross section.
They exhibit systematic changes in the same way in 
all the three systems.
As $K_0$ increases, the height $V_b$ and the curvature $\hbar \omega$ 
tend to increase and the barrier position $R_b$ decreases.
The same dependence on $K_0$ is obtained with Skyrme  
models that have different $K_0$ values \cite{prc2016}.
Systematic evolution of the barrier properties will also have a 
systematic effect on the fusion cross section.

\begin{table}[t]
\caption{\label{tab:16O208Pb}
Low-lying collective states of $^{16}$O, $^{40}$Ca, 
$^{48}$Ca and
$^{208}\mathrm{Pb}$, which are included in the coupled-channels calculations. The
excitation energies and electromagnetic transition strengths of $^{40}$Ca 
and $^{48}$Ca are taken from Refs.~\cite{mont12} and ~\cite{esbensen10},
and those of $^{16}$O and $^{208}$Pb are from Ref.~\cite{mort10}.}
\begin{center}
\begin{ruledtabular}
\begin{tabular}{cccccc}
Nucleus & $\lambda^\pi$ & $E^{*}$ (MeV) & $B(E\lambda)\uparrow$ & $\beta_{\lambda\,C}$ & $\beta_{\lambda\,N}$ \\
\hline
$^{16}\mathrm{O}$  & $3^-_1$ & 6.129 & 0.0015 $e^2\,b^3$ & 0.733 & 0.733\\
\hline
$^{40}$Ca & $2^+_1$ & 3.905 & 2.26 $\pm$ 0.14 W.u & 0.143 & 0.130\\
        & $3^-_1$ & 3.737 & 27 $\pm$ 4 W.u& 0.482 & 0.326\\ \hline
 $^{48}$Ca & $2^+_1$ & 3.832 & 1.71 $\pm$ 0.09 W.u & 0.123 & 0.185\\
        & $3^-_1$ & 4.507 & 5.0 $\pm$ 0.8 W.u& 0.244 & 0.185\\ \hline
$^{208}\mathrm{Pb}$ & $3^-_1$ & 2.615 & 0.611 $e^2\,b^3$ & 0.161 & 0.161\\
                     & $5^-_1$ & 3.198 & $\delta=0.35~\mathrm{fm}$ & 0.056 & 0.056\\
\end{tabular}
\end{ruledtabular}
\end{center}
\end{table}

The contribution of the excited states is essential in the fusion process\cite{dasgupta1998,balantekin1998,hagino2012,hagino2022,montagnoli2017,montagnoli2023}.
Table \ref{tab:16O208Pb} summarizes the excitation energies and the transition
strengths of the excited states to be taken into account in calculating the fusion cross sections.
In Tab.\,\ref{tab:16O208Pb}, $\beta_C$ are determined 
from the measured $B(E\lambda)$ values of each nucleus.
For $\beta_N$, the values are assumed equal to the values
of $\beta_C$ for ${}^{16}$O and ${}^{208}$Pb,
while they are extracted from the analysis of experimental data on elastic and inelastic scattering~\cite{mont12} 
for $^{40}$Ca and $^{48}$Ca.
These collective excitations provide the leading coupling effects,
whereas the influence of high-lying states is negligible in the fusion.
Coupled-channel calculations are performed with the {\tt CCFULL} code \cite{ccfull} using
Woods-Saxon representations of the nuclear potential derived from the
frozen-density EDM.

\begin{figure}[t]
  \centering
\includegraphics[width=0.5\textwidth]{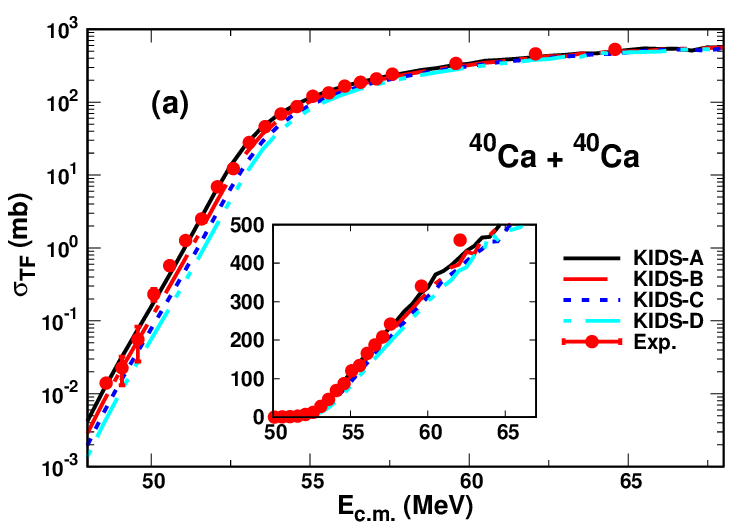}
\includegraphics[width=0.5\textwidth]{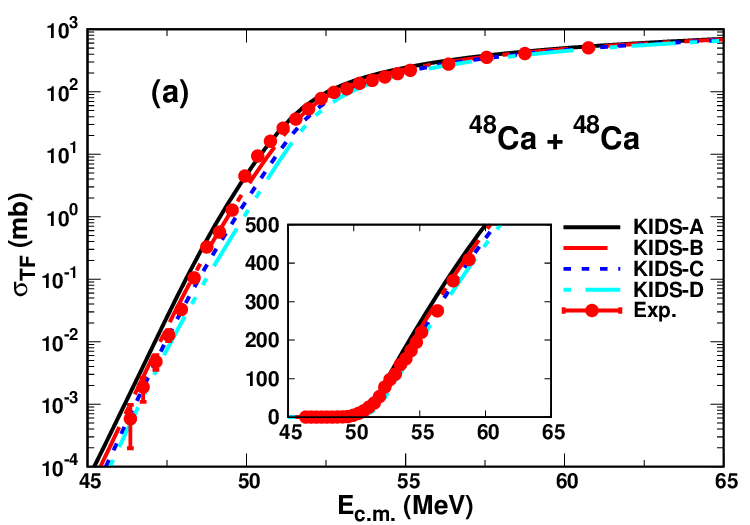}
\includegraphics[width=0.5\textwidth]{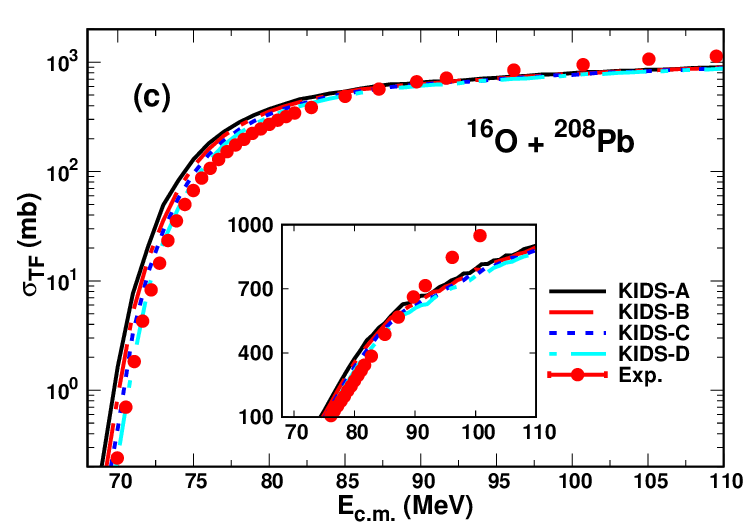}
\caption{
    Fusion cross sections calculated with {\tt CCFULL} with KIDS-A, B, C and D models for
    (a) $^{40}\mathrm{Ca}+{}^{40}\mathrm{Ca}$, 
    (b) $^{48}\mathrm{Ca}+{}^{48}\mathrm{Ca}$ and
    (c) $^{16}\mathrm{O}+{}^{208}\mathrm{Pb}$. 
    The experimental data shown by
    red circles are taken from Refs. \cite{mont12,Stefanini09,mort10}. 
  }
  \label{fig:results-cross-sections}
\end{figure}

Figure~\ref{fig:results-cross-sections} compares the calculated fusion
cross sections with the experimental data for 
(a) $^{40}\mathrm{Ca}+{}^{40}\mathrm{Ca}$, 
(b) $^{48}$Ca+$^{48}$Ca
and (c) $^{16}\mathrm{O}+{}^{208}\mathrm{Pb}$. 
For the $^{40}\mathrm{Ca}+{}^{40}\mathrm{Ca}$ system, four EDFs reproduce 
the measured fusion cross sections reasonably well over the full energy range 
as shown in Fig.~\ref{fig:results-cross-sections} (a).
Their predictions begin to separate below the Coulomb barrier. 
The differences are not significant, but the KIDS-A model agrees best with 
the data among the four models
and the KIDS-D model shows the strongest suppression.
This behavior is consistent with the potentials shown in
Fig.~\ref{fig:results-potentials} (a)
and the characteristics of the potential barrier in Tab.~\ref{tab:ws-fit-parameters}. 
Weaker attraction pushes the barrier upward and reduces its penetrability. 
Because sub-barrier fusion proceeds predominantly through quantum tunneling, 
small variations in the barrier region are enhanced
in the calculated cross sections. 
The differences among the EDF
predictions therefore reflect the EoS dependence of the barrier properties.
Above the barrier, the differences decrease with higher energies and 
curves eventually converge to a single line.

For the $^{48}$Ca+$^{48}$Ca system,
similar to $^{40}$Ca+$^{40}$Ca,
theoretical result agrees well with data below and above
the Coulomb barrier.
However, contrary to $^{40}$Ca+$^{40}$Ca where KIDS-A model
shows consistent agreement with data in the sub-barrier region,
KIDS-B tends to reproduce the data better than KIDS-A
in the interval $48 \lesssim E_{\rm cm} \lesssim 50$ MeV,
and the agreement is improved with KIDS-C at the energies
below 48 MeV.
Density distribution of $^{48}$Ca is controlled by both incompressibility
and symmetry energy.
Therefore, the result of $^{48}$Ca+$^{48}$Ca can be interpreted that
the combinations of $K_0=240 \sim 250$ MeV, $J=31\sim32$ MeV
and $L\sim 58$ MeV are best to describe the sub-barrier fusion
processes in $^{48}$Ca+$^{48}$Ca.
These ranges are consistent with the ones constrained by the neutron
star data \cite{npsm2022}.

For the $^{16}\mathrm{O}+{}^{208}\mathrm{Pb}$ system, 
KIDS-D provides the closest agreement with the data
in the sub-barrier domain,
whereas KIDS-A, KIDS-B and KIDS-C overestimate the data. 
The ordering of the calculated cross sections follows that of the
potentials shown in Fig.~\ref{fig:results-potentials} (c). 
The higher barriers in the KIDS-D model reduces penetration
probability and suppress the low-energy cross sections. 
Even sub-MeV changes in the potential can become observable in the fusion
cross section because of the
exponential sensitivity of tunneling to the barrier parameters. 
At energies above the barrier,
theoretical result agrees with the experimental data accurately in $^{40}$Ca+$^{40}$Ca
and $^{48}$Ca+$^{48}$Ca, but the theoretical result 
underestimates the data in $^{16}$O+$^{208}$Pb.
Because $^{16}$O+$^{208}$Pb is a much bigger system than $^{40}$Ca+$^{40}$Ca
and $^{48}$Ca+$^{48}$Ca,
frozen-density approximation could be a too simplified assumption for the system,
and additional effects such as Pauli blocking and deformation in the composite
system may play a role.


\subsection{Barrier distributions}

\begin{figure}[t]
  \centering
\includegraphics[width=0.5\textwidth]{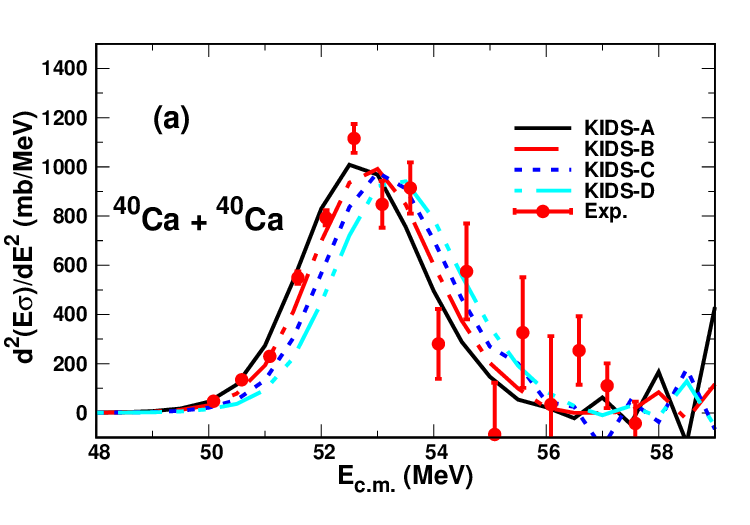}
\includegraphics[width=0.5\textwidth]{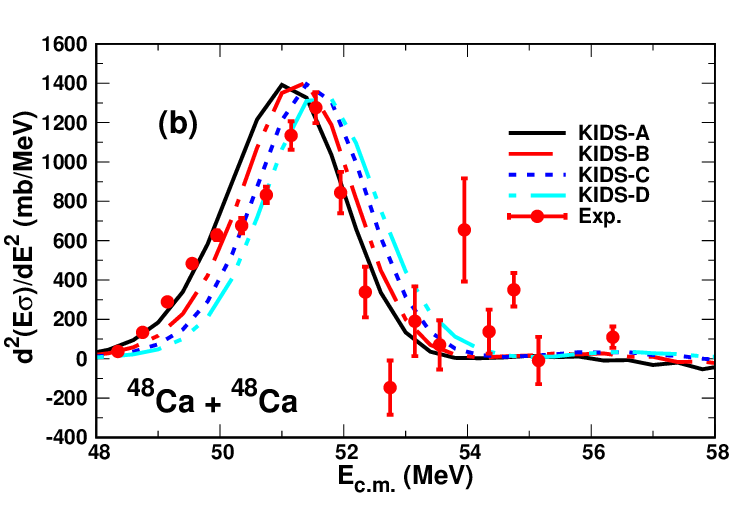}
\includegraphics[width=0.5\textwidth]{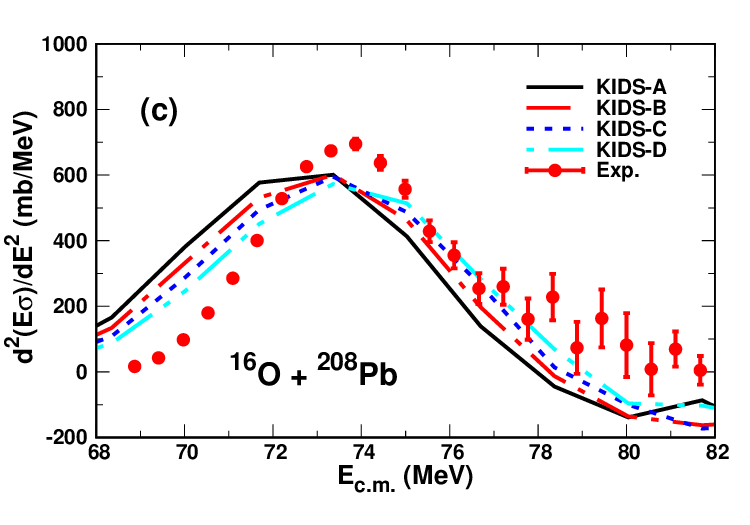}
  \caption{
    Fusion barrier distributions for
    (a) $^{40}\mathrm{Ca}+{}^{40}\mathrm{Ca}$, 
    (b) $^{48}\mathrm{Ca}+{}^{48}\mathrm{Ca}$
    and
    (c) $^{16}\mathrm{O}+{}^{208}\mathrm{Pb}$, 
    obtained with KIDS-A, KIDS-B, KIDS-C, and KIDS-D as indicated in the legends.
    The experimental data 
    shown by red circles with error bars are taken from Refs. \cite{mont12,Stefanini09,mort10}.
  }
  \label{fig:results-barrier-distributions}
\end{figure}

The fusion barrier distribution \cite{rowley1991,leigh1995} provides a differential probe of the reaction dynamics\cite{dasgupta1998,balantekin1998,hagino2012}. 
It represents the shape of 
the distribution of the
Coulomb barrier in the fusion absorption region
as a function of the incident energy.
It is defined as
\begin{equation}
  D(E) = \frac{d^2(E\sigma)}{dE^2},
\end{equation}
where $\sigma$ is the total fusion cross section. 
The calculated barrier distributions for the three systems are 
shown in Fig.~\ref{fig:results-barrier-distributions}.

Figure~\ref{fig:results-barrier-distributions} (a) displays the
distribution for $^{40}\mathrm{Ca}+{}^{40}\mathrm{Ca}$.
The calculated barrier
distributions exhibit a single dominant peak near the nominal fusion
barrier. 
The centroid follows the same ordering as the nuclear potentials: 
KIDS-A appears at the lowest energy and KIDS-D at the highest.
$E_{\rm cm}$ values at peaks are about 52.5 MeV for KIDS-A,
about 53 MeV for KIDS-B and KIDS-C and about 53.5 MeV for KIDS-D,
so the difference between the models is not significant.
The calculated widths are nearly unchanged. 
Thus, the principal EoS effect is a shift of the effective barrier rather than 
a substantial modification of its shape.

We present the barrier distribution for $^{48}$Ca+$^{48}$Ca in
Fig.~\ref{fig:results-barrier-distributions} (b).
Similar to $^{40}$Ca+$^{40}$Ca, the effect of the EoS appears as
shifts of the barrier distributions without significantly 
altering their shapes 
from the leftmost KIDS-A to the rightmost KIDS-D.
It is interesting to note that the peak in the barrier distribution with KIDS-A is clearly distinguished
from that with the other models,
but the positions of peaks are adjacent to each other
for KIDS-B, KIDS-C and KIDS-D models,
and they reproduce the experimental data accurately.
Similarity of the barrier distribution among KIDS-B, KIDS-C and KIDS-D is
consistent with the similarity of the models in the fusion cross seciton.

The barrier distribution of the $^{16}\mathrm{O}+{}^{208}\mathrm{Pb}$ system
is shown in Fig.~\ref{fig:results-barrier-distributions} (c).
Coordinates of the peaks are within 1 MeV gap along the 
horizontal axis and about 100 mb/MeV discrepancy in the height 
from the experimental peak point.
These differences are similar to those of $^{40}$Ca+$^{40}$Ca
and $^{48}$Ca+$^{48}$Ca. 
On the other hand, while the peaks are shifted
horizontally in the previous two Ca systems, 
the peak positions of the four models are almost identical in the
$^{16}$O+$^{208}$Pb system, but the shape of the barrier
depends on the model.
Different form of the dependence on the model might imply that
the adopted static EDFs and frozen-density EDM do not include all dynamical processes responsible for the observed barrier fragmentation
in the $^{16}$O+$^{208}$Pb system.

The barrier-distribution analysis therefore supports the conclusion that
the EoS dependence enters primarily through the nuclear interaction in
the surface-overlap region.
The ground-state densities vary only
slightly, but the associated changes in the barrier are amplified by
quantum tunneling and become observable in the sub-barrier fusion cross
sections.

\subsection{Effective mass}

\begin{table}[t]
\caption{Parameters for the KIDS functional in the form of Skyrme force. These are  
for the given isoscalar and isovector effective masses at the saturation density. Units of $t_1$, $y_1$, $t_2$, $y_2$, $t_{32}$, $y_{32}$ and $W_0$ are MeV fm$^5$. 
All the models have identical values of $t_0= -1772.04$ and $y_0= -127.52$ in MeV fm$^3$, 
$t_{31}= 12216.73$ and $y_{31}= -11969.99$ in MeV fm$^4$, and $t_{33}= 0$ and 
$y_{33}= -22955.28$ in MeV fm$^6$.}
\begin{center}
\begin{ruledtabular}
\begin{tabular}
{cccccccccc}
 & $t_1$ & $y_1$ & $t_2$ & $y_2$ & $t_{32}$ & \phantom{aa} $y_{32}$ \phantom{aa} & $W_0$ 
 & $\mu_{\rm IS}$ & $\mu_{\rm IV}$ \\ \hline
KIDS0-m*71 & 452.09 & $-674.87$ & $-149.88$ & 70.45 & $-2572.65$ & 45248.10 & 152.47 & 0.7 & 1.0 \\
KIDS0-m*81 & 376.97 & $-537.01$ & $-85.06$ & $-46.81$ & $-1233.17$ & 40208.67 & 140.93 & 0.8 & 1.0 \\
KIDS0-m*11 & 273.01 & $-367.16$ & 34.54 & $-247.93$ & 642.12 & 33153.48 & 125.25 & 1.0 & 1.0 \\
KIDS0-m*77 & 451.80 & $-400.40$ & 77.09 & $-213.05$ & $-2572.65$ & 37593.40 & 145.23 & 0.7 & 0.7 \\
KIDS0-m*78 & 451.92 & $-515.08$ & $-17.07$ & $-95.44$ & $-2572.65$ & 40782.86 & 148.29 & 0.7 & 0.8 \\
KIDS0-m*79 & 452.02 & $-603.52$ & $-91.29$ & $-2.73$ & $-2572.65$ & 43263.55 & 150.61 & 0.7 & 0.9 
\end{tabular}
\end{ruledtabular}
\end{center}

\label{tab:effmass}
\end{table}

In the KIDS model, and thus the Skyrme model, as one can see in Eqs. (\ref{eq:iseffmass}) and (\ref{eq:iveffmass}),
effective masses are determined by $t_1$, $y_1$, $t_2$ and $y_2$.
These parameters also contribute to the derivatives of the density 
[fourth line in Eq. (\ref{eq:kids-hamiltonian-density})], 
so a different effective mass can have an effect on the surface properties of a nucleus
and consequently potential barriers between two nuclei.
In the present work, we take KIDS0 functional \cite{prc2019a} as a baseline
EDF and implement the variation of effective masses on top of the model.
In the KIDS0 functional, $y_1=y_2=0$ is assumed for simplicity,
and the effective masses are obtained as results of the fitting.
If we assume non-zero values for $y_1$ and $y_2$, they can be calibrated to
produce specific values of the effective mass in the course of fitting to 
nuclear data.
Range of uncertainties of isoscalar and isovector effective masses is
$(0.7 \sim 1.0)m$, so we consider six combinations of $(\mu_{\rm IS}, \mu_{\rm IV})=$
(0.7, 1.0), (0.8, 1.0), (1.0, 1.0), (0.7, 0.7), (0.7, 0.8) and (0.7, 0.9).
The first three sets are chosen to show the effect of isoscalar effective mass,
and the last three sets for that of the isovector effective mass.
Each model is named KIDS0-m*71, m*81, m*11, m*77, m*78 and m*79.
The parameters of the model are summarized in Tab. \ref{tab:effmass}.

\begin{figure}[t]
  \centering
\includegraphics[width=0.5\textwidth]{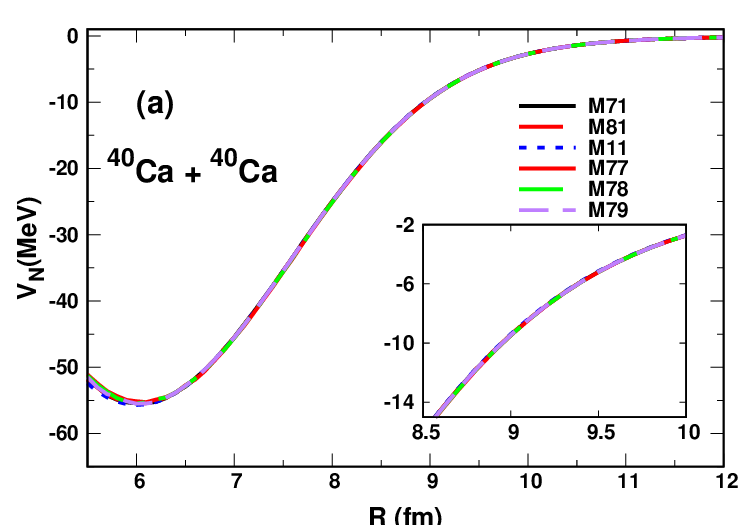}
\includegraphics[width=0.5\textwidth]{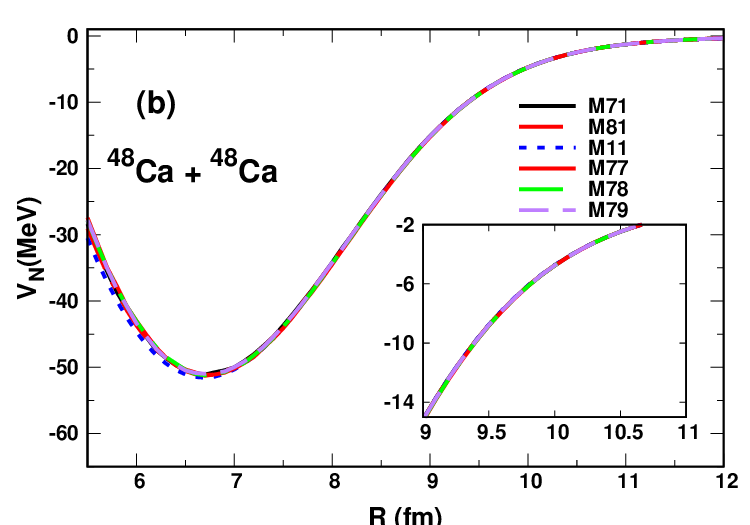}
\includegraphics[width=0.5\textwidth]{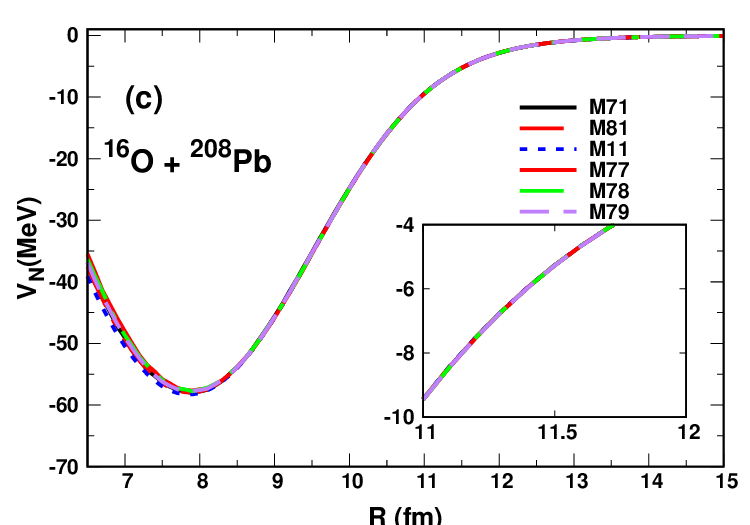}

  \caption{Nuclear potentials $V_N(R)$ for
    (a) $^{40}\mathrm{Ca}+{}^{40}\mathrm{Ca}$, 
    (b) $^{48}\mathrm{Ca}+{}^{48}\mathrm{Ca}$ and
    (c) $^{16}\mathrm{O}+{}^{208}\mathrm{Pb}$
    obtained from the frozen-density
    EDM potentials. The curves show the results for KIDS0-m*71, m*81, m*11,
    m*77, m*78 and m*79 with the line styles indicated in each panel.
    The inset in each panel enlarges the surface region relevant to the
    fusion barrier.}
  \label{fig:effctive-potential}
\end{figure}

Figure \ref{fig:effctive-potential} depicts the nucleus-nucleus potential calculated 
with EDM for the six KIDS0 based effective mass models, 
representing (a) $^{40}$Ca+$^{40}$Ca,
(b) $^{48}$Ca+$^{48}$Ca and 
(c) $^{16}$O+$^{208}$Pb.
Small differences are recognizable
inside the minimum of potential pocket, but in 
the regions around the potential barriers, 
even the magnified figures in the insets
can hardly distinguish differences among the models.
Table\,\ref{tab:eff-barrier} summarizes the characteristics
of the barriers at the peak for the six effective mass models.
The numbers in the table show the similarity of the models
quantitatively.
Maximum difference between the models is about 1.7\,\% for $\hbar \omega$
in $^{40}$Ca+$^{40}$Ca,
but the differences are below 0.5\,\% for the remaing results.
Unlike the initial expectation that the effective mass may have certain correlation
with density profiles in the surface, 
interaction potentials demonstrate that the nucleus-nucleus potential
around the barrier is strongly independent of both isoscalar and isovector
effective masses.

\begin{table}[t]
    \caption{
    The barrier quantities of potentials 
    obtained with different KIDS models. The quantities   $V_b$, $R_b$, and $\hbar\omega$  are the uncoupled barrier height, the radius, and the curvature in units of MeV, fm and MeV, respectively.
  }
  \begin{center}
  \begin{ruledtabular}
      
  \begin{tabular}{cccccccc}
    System & Quantity & m*71 & m*81 & m*11 & m*77 & m*78 & m*79 \\
    \hline
  $^{40}\mathrm{Ca}+{}^{40}\mathrm{Ca}$
      & $V_b$ (MeV) & 55.46 & 55.47 & 55.49 & 55.46 & 55.47 & 55.47 \\
      & $R_b$ (fm) & 9.51 & 9.51 & 9.50 & 9.52 & 9.51 & 9.51 \\
      & $\hbar\omega$ (MeV) & 3.47 & 3.47 & 3.48 & 3.53 & 3.53 & 3.53 \\ 
      \hline
    $^{48}\mathrm{Ca}+{}^{48}\mathrm{Ca}$
      & $V_b$ (MeV) & 52.86 & 52.87 & 52.87 & 52.85 & 52.85 & 52.86 \\
      & $R_b$ (fm) & 10.04 & 10.04 & 10.04 & 10.05 & 10.09 & 10.05 \\
      & $\hbar\omega$ (MeV) & 3.08 & 3.09 & 3.09 & 3.09 & 3.09 & 3.09 \\ 
      \hline
  
$^{16}\mathrm{O}+{}^{208}\mathrm{Pb}$
      & $V_b$ (MeV) & 76.92 & 76.94 & 76.93 & 76.92 & 76.93 & 76.92 \\
      & $R_b$ (fm) & 11.35 & 11.35 & 11.35 & 11.36 & 11.35 & 11.35 \\
      & $\hbar\omega$ (MeV) & 4.20 & 4.21 & 4.22 & 4.22 & 4.22 & 4.21 \\ 
  \end{tabular}
  \end{ruledtabular}
  \end{center}
  \label{tab:eff-barrier}
\end{table}

\begin{figure}[t]
  \centering
\includegraphics[width=0.5\textwidth]{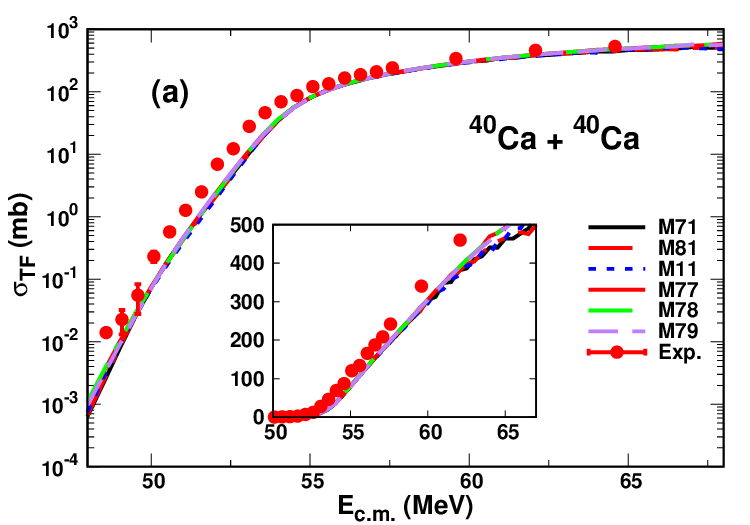}
\includegraphics[width=0.5\textwidth]{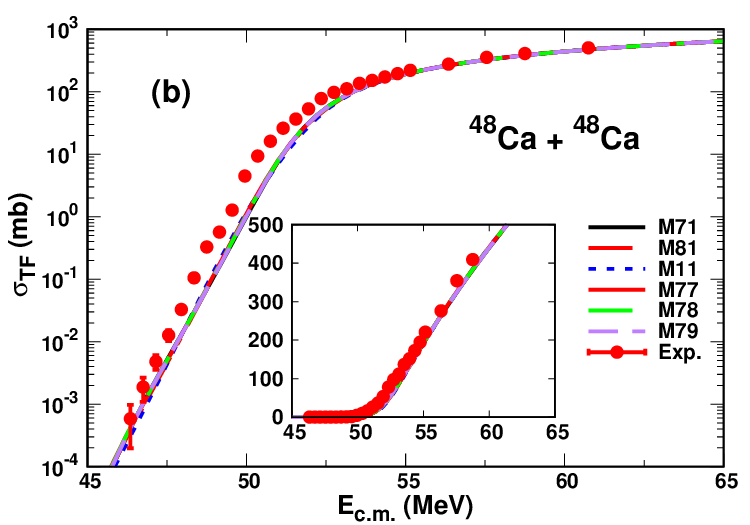}
\includegraphics[width=0.5\textwidth]{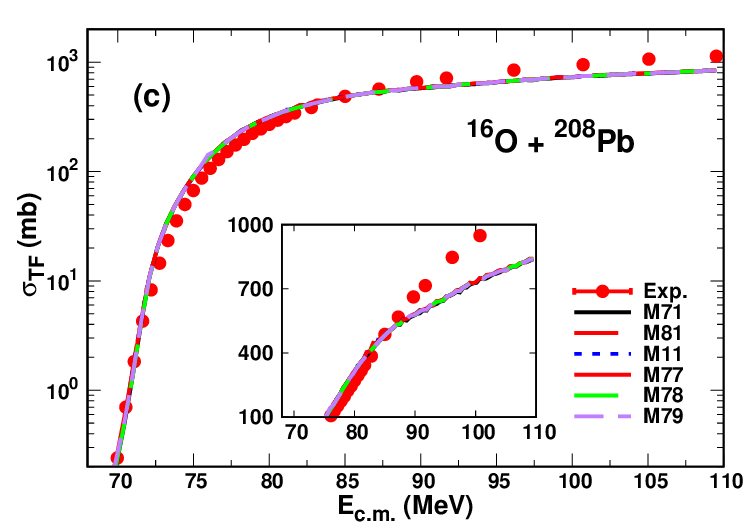}

  \caption{
    Fusion cross sections calculated with {\tt CCFULL} with KIDS0-m*71, m*81, m*11, m*77, m*78
    and m*79 models for
    (a) $^{40}\mathrm{Ca}+{}^{40}\mathrm{Ca}$, 
    (b) $^{48}\mathrm{Ca}+{}^{48}\mathrm{Ca}$ and
    (c) $^{16}\mathrm{O}+{}^{208}\mathrm{Pb}$. 
    The experimental data shown by
    red circles are taken from Refs. \cite{mont12,Stefanini09,mort10}. 
        }
  \label{fig:effctive-fusion}
\end{figure}

It is not difficult to imagine that the fusion cross sections of the 
six effective mass models may be similar to each other.
Figure \ref{fig:effctive-fusion} verifies the expectation and 
confirms negligible sensitivity to the effective masses.
It is noteworthy that the results of effective mass models
are closest to the result of KIDS-C for $^{40}$Ca+$^{40}$Ca,
and resemble most the results of KIDS-D
for $^{48}$Ca+$^{48}$Ca and $^{16}$O+$^{208}$Pb.
These similarities can be attributed to the nuclear matter behavior of the 
KIDS0 functional: $\rho_0=0.16$ fm$^{-3}$, $E_B=16.0$ MeV, $K_0=240$ MeV
for the symmetric nuclear matter and $J=32.8$ MeV and $L=49.1$ MeV for the
symmetry energy.
Incompressibility of the KIDS0 model is similar to the value of 
the KIDS-C model,
and the value of $L$ of KIDS0 is close to that of KIDS-D.
The result of effective mass models follows the same pattern
we obtained with the KIDS-A, B, C and D models, i.e. 
dependence on the EoS arises from
the incompressibility for the symmetric system and more substantial
dependence on the symmetry energy in the asymmetric system.

\section{Summary}

In the present work, the nucleus-nucleus interaction potential was calculated with the energy density method. 
The KIDS functional, which is one of the EDF models, was used to construct the interaction potential. 
We studied the influence of incompressibility,
symmetry energy and in-medium effective mass of nucleon 
on the nuclear reaction
through fusion cross section and barrier distribution.  

Firstly, the dependence on the EoS was investigated for the proton and the neutron densities in the ground-state of $^{16}$O, $^{40}$Ca,
$^{48}$Ca and $^{208}$Pb 
with the KIDS-A, KIDS-B, KIDS-C and KIDS-D EDF models. 
In the central region, the magnitude of the density was found to be in the order KIDS-A $<$ KIDS-B $<$ KIDS-C $<$ KIDS-D. 
Since $^{16}$O and $^{40}$Ca are symmetric nuclei, the difference between the models appears primarily due to $K_0$, 
while $^{48}$Ca and $^{208}$Pb are asymmetric nuclei, 
so not only the incompressibility, 
but also the symmetry energy should be considered.

Secondly, we calculated the interaction potential for $^{40}$Ca+$^{40}$Ca,
$^{48}$Ca+$^{48}$Ca and $^{16}$O+$^{208}$Pb systems 
with frozen-density EDM using 
the KIDS-A, KIDS-B, KIDS-C and KIDS-D EDF models.
As a result, we found that the model dependence appears manifestly around 
overlapping surfaces, that is, the potential 
is most attractive in the KIDS-A model and 
least attractive in the KIDS-D model. 

To calculate fusion cross sections, we fitted the interaction potential with a Woods-Saxon type to use the {\tt CCFULL} code.
In all the three systems, calculated sub-barrier fusion cross sections
agreed well with the experimental data.
Above the Coulomb barrier, cross sections were reproduced accurately
in $^{40}$Ca+$^{40}$Ca and $^{48}$Ca+$^{48}$Ca,
but theory underestimates the data in $^{16}$O+$^{208}$Pb.
Dependence on the EoS obtained in the internuclear potential 
was mapped onto the fusion cross section.
Difference between the models is not significant,
but for the $^{40}$Ca+$^{40}$Ca system, KIDS-A model agreed with the
data best among the four models.
In the systems with asymmetric nuclei,
KIDS-B and KIDS-C reproduced the data most closely
in the sub-barrier region in $^{48}$Ca+$^{48}$Ca
and KIDS-D was the best model in $^{16}$O+$^{208}$Pb.
From these results, one can deduce that the role of symmetry energy
becomes important as a nucleus is more asymmetric.

As another way to probe the influence of the EoS,
barrier distributions were calculated for the three systems and
the results were compared with the experimental data.
It was confirmed that a model that reproduces the fusion cross section
data most accurately also agrees best with the barrier distribution
data.
Regardless of the system, the best models have the positions of 
peaks in the barrier distribution within 1 MeV difference in $E_{\rm cm}$ and about 100 mb/MeV
difference in the height of the peaks from the data.

Finally, we studied the role of isoscalar and isovector effective masses within KIDS0 functional model on the interaction potential and fusion cross section for $^{40}$Ca+$^{40}$Ca,
$^{48}$Ca+$^{48}$Ca and $^{16}$O+$^{208}$Pb systems.
The influence of the effective masses was  
very small in both the interaction potential and the fusion cross sections. 

In conclusion, the effect of incompressibility 
and symmetry energy
appears apparently around overlapping surface region but the effective masses hardly affect these reactions. 
Therefore, low-energy heavy-ion reactions could serve as ways to constrain
the uncertainties of the nuclear matter EoS.
In the opposite way, by applying different EoS models to low-energy
reactions, one can estimate the uncertainties of the reaction
processes stemming from the nuclear matter EoS.
Good agreement with the data for $^{40}$Ca+$^{40}$Ca and $^{48}$Ca+$^{48}$Ca
encourages applications to more diverse systems whose masses are
similar to the two Ca systems.
Incosnsitency with data in the above-barrier region for  $^{16}$O+$^{208}$Pb evokes investigations on the origin of the
discrepancy and improvement of EDM so that it could be a general
method applicable to a broad range of masses.

\section*{Acknowledgements}
We thank Alberto Stefanini and 
Giovanna Montagnoli for providing us with the experimental data.
This work was supported by the National Research Foundation (grant nos. 2023R1A2C1003177,
RS-2023-00247042,
RS-2024-00460031,
RS-2025-16066382,
RS-2025-16070408,
RS-2026-25489753),
and by JSPS KAKENHI Grants Nos. 23K03414 and 26H01416.

\end{document}